# Charge-Based Compact Model for Bias-Dependent Variability of 1/f Noise in MOSFETs

Nikolaos Mavredakis[1], Nikolaos Makris[1], Predrag Habas[2], and Matthias Bucher[1], *Member, IEEE*

*Abstract*—Variability of low frequency noise (LFN) in MOSFETs is bias-dependent. Moderate- to large-sized transistors commonly used in analog/RF applications show 1/f-like noise spectra, resulting from the superposition of random telegraph noise (RTN). Carrier number and mobility fluctuations are considered as the main causes of low frequency noise. While their effect on the bias-dependence of LFN has been well investigated, the way these noise mechanisms contribute to the bias-dependence of variability of LFN has been less well understood. LFN variability has been shown to be maximized in weak inversion (sub-threshold), while increased drain bias also increases LFN variability. However, no compact model has been proposed to explain this bias-dependence in detail. In combination with the charge-based formulation of LFN, the present paper proposes a new model for bias-dependence of LFN variability. Comparison with experimental data from moderately-sized NMOS and PMOS transistors at all bias conditions provides insight into how carrier number and mobility fluctuation mechanisms impact the bias-dependence of LFN variability.

## I. INTRODUCTION

SCALING down of CMOS technologies leads to both shrinking of device area and operation in sub-threshold region. Under these conditions, variability of low frequency noise (LFN) becomes dominant and thus LFN becomes a key limiting factor in circuit design. In small devices, LFN is dominated by random telegraph signals (RTS) [1] which are caused by the capture and subsequent emission of charges at discrete trap levels near the oxide interface [2] - [7]. Each carrier trapped close to the silicon-oxide interface causes RTS in time domain, corresponding to a Lorentzian spectrum. The power spectral density (PSD) of LFN in a large-area MOSFET results from the superposition of such Lorentzians while the increased number of traps ensures the inversely proportional to frequency behavior (~ 1/f). This trapping - detrapping mechanism causes carrier number fluctuations (McWhorter model [8]), which is one of the main contributors to LFN. This effect is adequately covered by a number of basic LFN models available in bibliography [9] – [12].

On the other hand, transistors in analog and RF circuits typically have a very large area, from tens to thousands of square microns. The Lorentzian-like spectra are responsible for strong LFN deviation in small area devices where number of traps is quite low [5], [13]. In larger-area transistors, LFN variability is mainly affected by operating conditions and as it will be shown in this work, is connected with carrier number fluctuation ($\Delta N$) [8] and mobility fluctuation ($\Delta \mu$) [14], [15] effects. 1/f noise variability is minimum in high-current region [16], while it increases as inversion becomes lower reaching a plateau in weak inversion [17]. Finally, the non-uniformity of channel under high drain bias causes higher variability in saturation as compared to linear region [14]. Studies of statistical LFN variability have focused mainly on area-dependence [13], [18] – [22], while attempts have been made to describe the bias-dependence of noise variability [4], [7], [16]. A simple empirical model relating LFN variability to transconductance-to-current ratio $g_m/I_D$ [23] has been shown to provide satisfactory results for saturation from weak to strong inversion. However, no physics-based and truly compact model for the bias-dependence of LFN variability in MOSFETs has been proposed so far.

[1]N. Mavredakis, N. Makris and M. Bucher are with the School of Electronic and Computer Engineering, Technical University of Crete, Chania 73100, Greece (+30-2821037210, bucher@electronics.tuc.gr).
[2]P. Habas is with EM Microelectronic, Marin 2074, Switzerland.

The main goal of the present work is to propose a complete physics-based compact model of 1/f noise variability in all operating regions of the MOSFET. A charge-based model of LFN including $\Delta N$ and $\Delta \mu$ fluctuations has been described formerly [24] – [26]. In this work, we apply the mathematical equations of statistics to the above charge-based model of $\Delta N$ and $\Delta \mu$ fluctuations. The resulting compact model provides new insight into bias-dependence of LFN variability of MOSFETs. The model is shown to agree with measurements in N- and P-channel MOSFETs in all bias conditions, and results are in general agreement with known literature. The model for noise variability can easily be implemented in the charge-based EKV3 compact MOSFET model [25], [27], [28].

## II. DEVICES AND MEASUREMENTS

On-wafer 1/f noise measurements were performed on N- and P-channel devices in an experimental 180nm CMOS process flow. LFN spectra of 30 dies of NMOS and PMOS W/L=5μm/2μm transistors are measured over one wafer in both saturation and linear regimes, with $|V_{DS}|$ = 1.2V and 50mV, respectively, from weak to strong inversion, with $|V_{GS}|$ = 0.35, 0.4, 0.45, 0.5, 0.55, 0.6, 0.8, 1.2, 1.8V. Additionally, drain bias dependence is analyzed, with $|V_{DS}|$ = 0.05, 0.15, 0.3, 0.6, 1.2V at two gate voltage values, $|V_{GS}|$ = 0.6, 1.2V, to closely examine the bias-dependence of 1/f

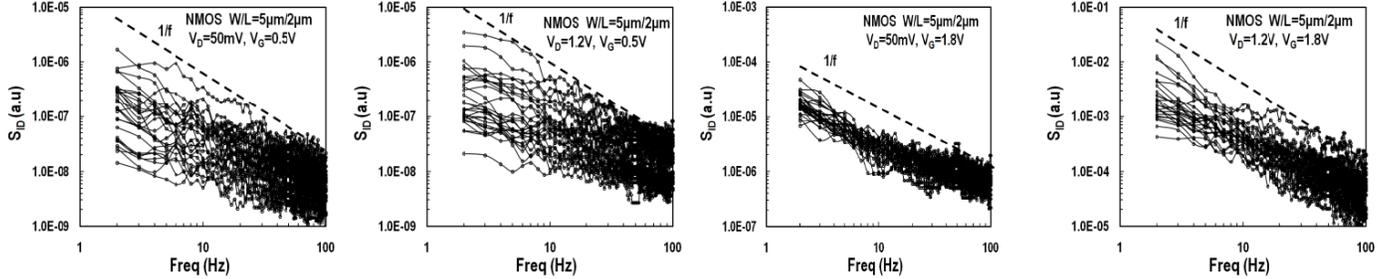

Fig. 1. Relative power spectral density of drain current noise $S_{ID}$ for N-channel MOSFETs with W=5μm, L=2μm. Bias conditions are around-threshold with $V_{GS}$=0.5V (leftmost: $V_{DS}$=50mV, left-center: $V_{DS}$=1.2V), and strong inversion with $V_{GS}$=1.8V (right-center: $V_{DS}$=50mV, rightmost: $V_{DS}$=1.2V). Variability is increased at low gate bias, and increases with increased drain bias.

noise variability versus bias conditions. The measured frequency range is from 2 Hz up to 2 kHz.

Fig. 1 shows the measured spectra of W/L=5μm/2μm transistors in different cases with a slope close to 1/f. Variability of 1/f noise is clearly seen to be maximized at low gate bias, and at high drain bias. Similar behavior can be observed in PMOS devices.

Fig. 2 shows the noise data, as gathered in the first analysis, averaged in a bandwidth of 10 - 50 Hz, referred to 1 Hz, showing measured output noise PSD $WLS_{ID}/I_D^2$ versus $I_D/(W/L)$, for N-channel and P-channel devices from subthreshold to strong inversion, in both linear and saturation modes.

## III. PHYSICAL 1/F NOISE MODEL

As mentioned above, the basic contributors for LFN are ΔN and Δμ effects. A charge-based model for the mean value of LFN covering these effects has been proposed [24] – [26]. Firstly, the normalized PSD $WLfS_{\delta i}/I_D^2$ of a local noise source, concerning a slice $\Delta x$, is given by [25] for both $\Delta N$ and $\Delta \mu$ effects, respectively:

$$\left.\frac{S_{\delta i_D^2}}{I_D^2}\right|_{\Delta N} WLf = \left(\frac{1}{q_i + 1/2} + \alpha \mu\right)^2 \frac{N_{tr}}{W \Delta x N_{SPEC}^2} \tag{1}$$

$$\left.\frac{S_{\delta i_D^2}}{I_D^2}\right|_{\Delta \mu} WLf = \frac{1}{q_i} \frac{\alpha_H WL}{W \Delta x N_{SPEC}} \tag{2}$$

where $q_i$ is the local inversion charge density, $N_{tr}=WLkT\lambda N_T$ is the number of traps, with $N_T$ the oxide volumetric trap density per unit energy in $eV^{-1}cm^{-3}$, evaluated close to the Fermi energy level, $k$ the Boltzmann constant, $T$ the absolute temperature, $q$ the electron charge, $f$ the frequency, $\lambda \approx 0.1$ nm the tunneling attenuation distance, $\alpha = \alpha c 2n U_T C_{ox}$ is related to the Coulomb scattering, $\mu$ is the carrier mobility, $\alpha_H$ is the unitless Hooge parameter, and $N_{SPEC}$ the specific density in $cm^{-2}$ defined in Table I, and W, L the device width and length. The latter defines also essential charge-voltage and current-charge relationships in the charge-based model, as well as other quantities for normalization of current, voltage, and charges. The normalized inversion charge densities $q_{s(d)}$ at source and drain, respectively, govern all other model quantities, such as transconductances, transcapacitances and noise, at all bias conditions [25], [28].

### TABLE I
DEFINITIONS OF ESSENTIAL CHARGE-BASED MODEL QUANTITIES

| | | | |
|---|---|---|---|
| Pinch-off voltage | $V_P \cong (V_G - V_{TO})/n$ | Slope factor | $n = [\partial V_P/\partial V_G]^{-1}$ |
| Thermal voltage | $U_T = kT/q$ | Specific charge | $Q_{SPEC} = 2nU_TC_{ox}$ |
| Normalized voltages | $v_x = V_x|_{x=P,G(S),D}/U_T$ | Normalized charges | $q_{x(d)} = Q|_{x=Q(S)}/Q_{SPEC}$ |
| Specific current | $I_{SPEC} = 2nU_T^2\mu C_{ox}W/L$ | Specific density | $N_{SPEC} = 2kTnC_{ox}/q^2$ |
| Drain current | $i_d = I_{D(f)}/I_{SPEC} = i_f - i_r$ | | |
| Normalized current | $i_{f(r)} = q_{s(d)}^2 + q_{s(d)}$ | Derivative | $\frac{\partial q_i}{\partial v} = \frac{q_i}{2q_i + 1}$ |
| Charge-voltage relationship | $v_p - v_{ch} = 2q_i + \ln q_i$ | quantities | $\frac{\partial i}{\partial v} = q_i$ |

### TABLE II
NOISE MODEL PARAMETERS

| PARAMETER | SYMBOL | UNITS | NMOS | PMOS |
|---|---|---|---|---|
| NT | $N_T$ | $eV^{-1}cm^{-3}$ | $3.10^{16}$ | $1.10^{17}$ |
| AC | $\alpha_C$ | $VsC^{-1}$ | $7.10^3$ | $1.3.10^5$ |
| AH | $\alpha_H$ | - | $1.10^{-6}$ | $2.10^{-6}$ |
| - | $\alpha\mu$ | - | 0.168 | 0.486 |

After integrating (1) and (2) along the channel, the basic expressions for PSD of noise current are:

$$\left.\frac{S_{\Delta i_D^2}}{I_D^2}\right|_{\Delta N} = \frac{q^4 N_T \lambda}{kTWLn^2 C_{ox}^2 f} \left[\frac{1}{2i_d}\ln\left(\frac{2q_s+1}{2q_d+1}\right) + \frac{\alpha\mu}{1+q_s+q_d} + \left(\frac{\alpha\mu}{2}\right)^2\right] \tag{3}$$

$$\left.\frac{S_{\Delta i_D^2}}{I_D^2}\right|_{\Delta \mu} = \frac{\alpha_H q^2}{kTWLnC_{ox}f} \cdot \frac{1}{(1+q_s+q_d)}\left[1+\frac{\ln(q_s/q_d)}{2(q_s-q_d)}\right] \tag{4}$$

$$\frac{S_{\Delta I_D^2}}{I_D^2} = \frac{S_{\Delta I_D^2}}{I_D^2}\bigg|_{\Delta N} + \frac{S_{\Delta I_D^2}}{I_D^2}\bigg|_{\Delta \mu}$$

(5)

The above model (5) shows a good qualitative fit to the ln-mean data shown in Fig. 3, with parameters as in Table II. The latter are comparable to results in [27]. The mean (expected) value $E(WL_f S_{ID}/I_D^2)$ vs. $I_D/(W/L)$ is shown in linear and saturation regions for both NMOS and PMOS devices. The noise model is shown to be consistent with measurements. The noise model is also shown as lines in Fig. 2, where average noise (markers) represent the ln-mean data of the measurements.

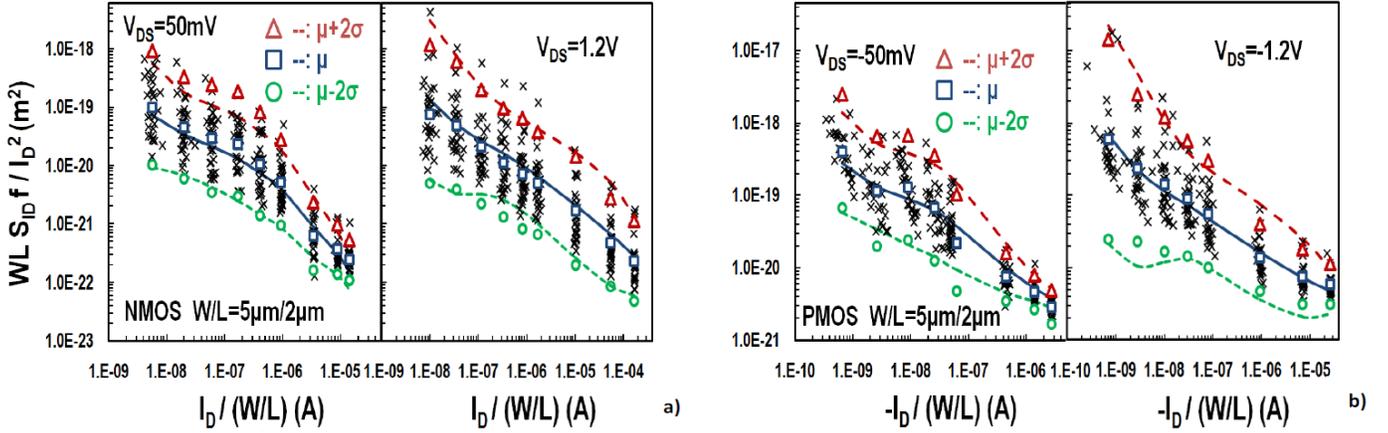

Fig. 2. Output noise $WLS_{ID}/I_D^2$ referred to 1Hz, vs. drain current $I_D/(W/L)$, measured for a) N-channel and b) P-channel devices with geometry W=5μm, L=2μm in linear region at $|V_{DS}| = 50$mV (left subplot) and saturation at $|V_{DS}| = 1.2$V (right subplot). Measured noise: crosses. Measured ln-mean noise, ±2-sigma deviation: open markers. EKV3 model: average noise (lines), ±2-sigma deviation (dashed).

While both $\Delta N$ and $\Delta \mu$ effects are technology-dependent, the $\Delta N$ effect appears to be always present while the appearance of $\Delta \mu$ effect depends on process [15]. As it can be seen from the same graphs, the $\Delta N$ effect determines noise level from moderate to strong inversion at high current, while correlated $\Delta N$-$\Delta \mu$ noise ($\alpha \mu$ effect) is also apparent. $\Delta \mu$ effect is dominant in sub-threshold region and appears as an increase over the weak inversion plateau of $WLS_{ID}/I_D^2$ as obtained from the $\Delta N$ effect.

As it will be shown in the statistical noise analysis that follows, these fundamental effects influence the 1/f noise statistics in an analogous way.

## IV. STATISTICAL 1/f NOISE MODEL

The noise of a MOS transistor itself is in fact the standard deviation of drain current. In order to model correctly the variations of 1/f noise, the parameters that are sensitive to these variations must be identified. From the physical LFN model, it can be concluded that $\Delta N$ and $\Delta \mu$ effects contribute to 1/f noise variability through variation of trap density $N_T$ and of parameter $\alpha_H$ in (3) and (4).

The procedure that will be followed is to calculate the variance of total noise PSD for each of $\Delta N$ and $\Delta \mu$ effects. This must take place before integration along the channel; otherwise, each $\Delta I_D$ caused by any fluctuation (e.g. a specific trap) would have the same effect, which is not valid [13].

Local noise sources in our modeling approach [25] are considered uncorrelated. The variance that will be included into the integral as shown below represents the local deviation corresponding to an elementary slice $\Delta x$ of the channel. Finally, by integrating from source to drain, the total variance is obtained by summing all the local contributions. According to basic statistics, $Var(f(y)) = [\delta f/\delta y]^2 \sigma_y^2$. The deviated parameters are the number of traps $N_{tr}$ for $\Delta N$, and $\alpha_H$ for $\Delta \mu$ effects, respectively.

### A. Variance of LFN due to Number Fluctuations

The next step will be to calculate the variance of the parameter deviated locally in the channel. The number of traps $N_{tr} = WLN_t$ follows a Poisson distribution [7], [16] and hence $\sigma^2_{Ntr} = WLN_t$, with $N_t = kT\lambda N_T$ [1], [3] the trap density in cm$^{-2}$.

The total normalized output noise due to $\Delta N$ effect is obtained by integrating (1) along the channel,

$$\frac{S_{\Delta I_D^2}}{I_D^2}\bigg|_{\Delta N} fWL = \frac{1}{LWN_{SPEC}^2}\int_0^1\left(\frac{1}{q_i + 1/2} + \alpha\mu\right)^2 N_t d\xi$$

(6)

where the integration variable was changed to $\xi = x/L$. Now, we calculate the variance of (6):

$$Var\left(\frac{S_{\Delta I_D^2}}{I_D^2}\bigg|_{\Delta N} fWL\right) = \left(\frac{1}{LWN_{SPEC}^2}\right)^2\int_0^1 Var\left[\left(\frac{1}{q_i + 1/2} + \alpha\mu\right)^2 N_{tr}\right]d\xi$$

(7)

where we have used $Var(ax) = a^2 Var(x)$. The variance in (7) now appears within the integral since local noise sources are uncorrelated and thus $Var(\Sigma x_i) = \Sigma(Var(x_i))$ where $x_i$ is the local noise source. To calculate variance due to number of traps, the partial derivative of the integrand of (7) with respect to number of traps $N_{tr}$ should be calculated:

$$Var\left[\left(\frac{1}{q_i+1/2}+\alpha\mu\right)^2 N_{tr}\right] =$$
$$(8)$$
$$= \left[\partial\left[\left(\frac{1}{q_i+1/2}+\alpha\mu\right)^2 N_{tr}\right]/\partial N_{tr}\right]^2 \sigma_{N_{tr}}^2 = \left(\frac{1}{q_i+1/2}+\alpha\mu\right)^4 WLN_t$$

By substituting (8) into (7) and by changing the integration variable $d\xi$ to $dq_i = -i_d/(1+2q_i)\,d\xi$ we obtain:

$$Var\left(\frac{S_{\Delta I_D^2}}{I_D^2}\bigg|_{\Delta N} fWL\right) = \frac{N_t}{N_{SPEC}^4 WL}\Lambda_D\bigg|_{\Delta N}$$
$$(9)$$

where,

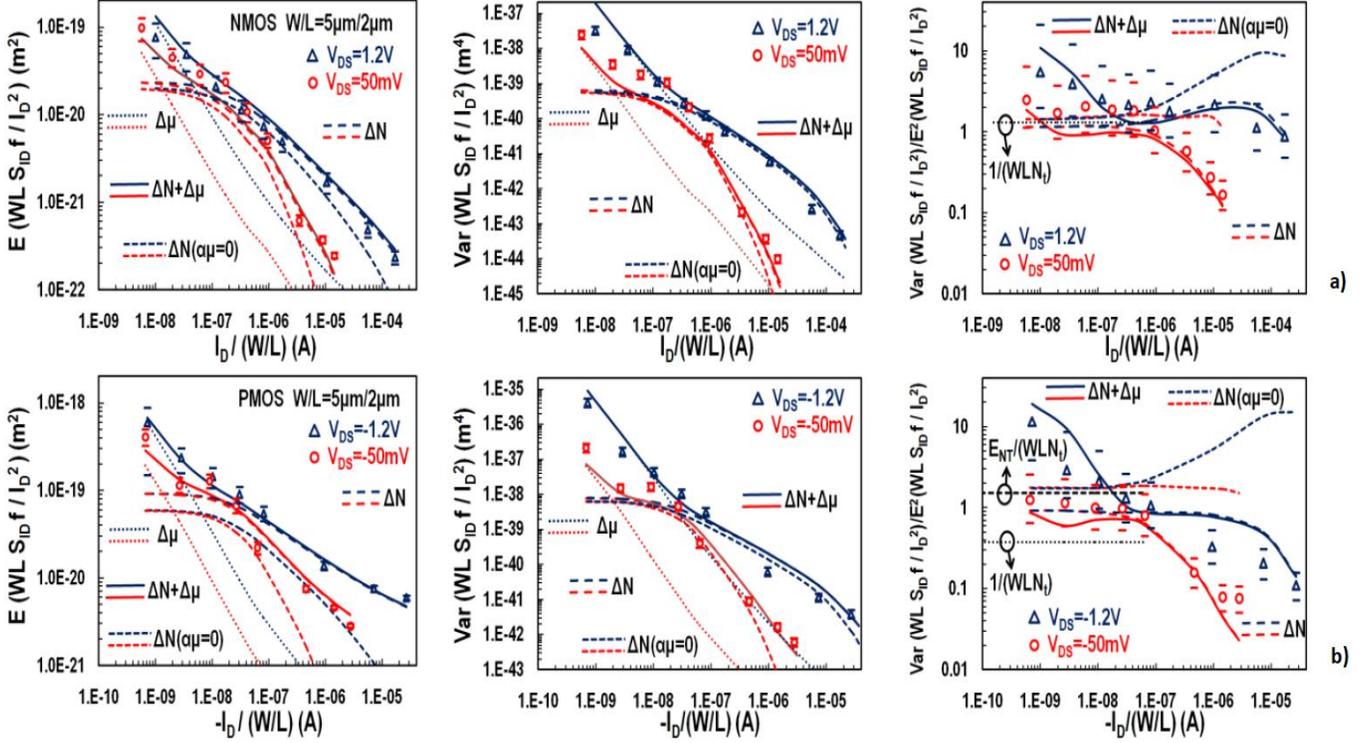

Fig. 3. Mean value (left), variance (center) and normalized variance (right) of output noise $WLS_{ID}/I_D^2$, referred to 1 Hz, vs. drain current $I_D/(W/L)$ for both N-channel (a) and P-channel (b) MOSFETs with W=5μm, L=2μm at |V_{DS}| = 50mV and 1.2V, respectively. Markers: measured data (with error bars indicating standard error), lines: entire model (ΔN + Δμ), dashed: individual contributions (ΔN, Δμ) to 1/f noise.

$$\Lambda_D\bigg|_{\Delta N} = \left(\begin{array}{l}\dfrac{1}{\left(q_s+1/2\right)^2\left(q_d+1/2\right)^2}+(\alpha\mu)^4+\dfrac{8(\alpha\mu)^3}{1+q_s+q_d}+\\[2mm]\dfrac{12(\alpha\mu)^2}{i_s}\ln\left(\dfrac{q_s+1/2}{q_d+1/2}\right)+\dfrac{8(\alpha\mu)}{(q_s+1/2)(q_d+1/2)(1+q_s+q_d)}\end{array}\right)$$
$$(10)$$

By transforming (3) as:

$$\frac{S_{\Delta I_D^2}}{I_D^2}\bigg|_{\Delta N} fWL = \frac{4N_t}{N_{SPEC}^2}\cdot K_D\big|_{\Delta N}$$
$$(11)$$

with:

$$K_D\big|_{\Delta N} = \frac{1}{2i_d}\ln\left(\frac{q_s+1/2}{q_d+1/2}\right)+\frac{\alpha\mu}{1+q_s+q_d}+\left(\frac{\alpha\mu}{2}\right)^2$$
$$(12)$$

we obtain the normalized variance of output noise due to ΔN effect, using (9) and (11), as:

$$\frac{K}{WL} = \frac{Var\left(WL\cdot S_{\Delta I_D^2}/I_D^2\right)}{E^2\left(WL\cdot S_{\Delta I_D^2}/I_D^2\right)} = \frac{1}{16\cdot WLN_t}\frac{\Lambda_D\big|_{\Delta N}}{\left(K_D\big|_{\Delta N}\right)^2}$$
$$(13)$$

where $K$ corresponds to the same quantity defined in [13]. Note that here, (13) provides an explicit model for $K$, in contrast to [13] where $K$ is defined as an integral expression which is not suitable for compact modeling. The term $\Lambda_D|_{\Delta N}/[16\,(K_D|_{\Delta N})^2]$ approaches unity in weak to moderate inversion irrespectively of $V_{DS}$, while it approaches unity throughout weak to strong inversion at small $V_{DS}$. (13) corresponds to the normalized variance of 1/f noise at all bias conditions, based on the assumption of a Poisson distributed number of traps of $N_{tr}=WLN_t$, as described above, assuming that only $\Delta N$ effect is present.

TABLE III
PARAMETERS OF LFN VARIABILITY MODEL

| PARAMETER | SYMBOL | UNITS | NMOS | PMOS |
|---|---|---|---|---|
| ENT | $E_{NT}$ | - | 1 | 4 |
| EAC | $E_{ac}$ | - | 0.3 | 0.25 |
| EAH | $E_{aH}$ | - | 0.1 | 0.1 |

### B. Variance of LFN due to Mobility Fluctuations

In the following, the variance due to $\Delta\mu$ effect – because of fluctuation of $\alpha_H$ parameter – is calculated. The total normalized output noise due to $\Delta\mu$ effect is obtained by integrating (2) along the channel:

$$\frac{S_{\Delta I_D^2}}{I_D^2}\bigg|_{\Delta\mu} fWL = \frac{1}{N_{SPEC}}\int_0^1 \frac{1}{q_i}\alpha_H d\xi \tag{14}$$

No information is a priori available on the standard deviation of $\alpha_H$ parameter. We can assume here that $\sigma^2_{\alpha H}=\alpha_H/(WLN_{SPEC})$, so that variability due $\Delta N$ and $\Delta\mu$ effects will have similar geometrical scaling.

By substituting (16) into (15) and by changing the integration variable from $d\xi$ to $dq_i$, we finally obtain:

$$Var\left(\frac{S_{\Delta I_D^2}}{I_D^2}\bigg|_{\Delta\mu}\right)WLf = \frac{\alpha_H}{WLN_{SPEC}^3}\frac{1}{i_d}\left(2\ln\frac{q_s}{q_d}+\frac{q_s-q_d}{q_s q_d}\right) \tag{17}$$

### C. Total Variance of LFN

The total variation of $1/f$ noise, according to (5), can finally be calculated as,

$$Var\left(WLf\frac{S_{\Delta I_D^2}}{I_D^2}\right) = Var\left(WLf\frac{S_{\Delta I_D^2}}{I_D^2}\bigg|_{\Delta N}\right) + Var\left(WLf\frac{S_{\Delta I_D^2}}{I_D^2}\bigg|_{\Delta\mu}\right) \tag{18}$$

We now calculate the variance of (14):

$$Var\left(\frac{S_{\Delta I_D^2}}{I_D^2}\bigg|_{\Delta\mu}fWL\right) = \left(\frac{1}{N_{SPEC}}\right)^2\int_0^1 Var\left[\frac{1}{q_i}\alpha_H\right]d\xi \tag{15}$$

The variance of the quantity $[(1/q_i)\alpha_H]$ is calculated as:

$$Var\left[\frac{1}{q_i}\alpha_H\right] = \left[\partial\left[\left(\frac{1}{q_i}\right)\alpha_H\right]/\partial\alpha_H\right]^2\sigma^2_{\alpha_H} = \left(\frac{1}{q_i}\right)^2\sigma^2_{\alpha_H} \tag{16}$$

Equations (9)-(10), (17) and (18) describe the new charge-based statistical 1/f noise compact model. The new model provides an explicit and truly compact formulation of bias-dependent LFN variability, formulated as a function of inversion charge densities at source and drain, $q_s$ and $q_d$. This model reflects the situation in long-channel devices and does not explicitly include short-channel effects.

The parameters of the physical model, $N_T$, $\alpha_C$, $\alpha_H$, presented in Table II, are also used in the statistical model. However, to provide the necessary flexibility to the statistical model, the following parameters are defined: $E_{NT}\approx N_{TS}/N_T$, $E_{aC}\approx\alpha_{CS}/\alpha_C$ and $E_{aH}\approx\alpha_{HS}/\alpha_H$ where $N_{TS}$, $\alpha_{CS}$, $\alpha_{HS}$ replace $N_T$, $\alpha_C$, $\alpha_H$ in (9) and (17). Furthermore, the term $1/(WLN_t)$ in (13) transforms to $E_{NT}/(WLN_t)$. The parameters of the statistical model, $E_{NT}$, $E_{aC}$ and $E_{aH}$, shown in Table III, are obtained by fitting (18) to measured variance data, as will be discussed below.

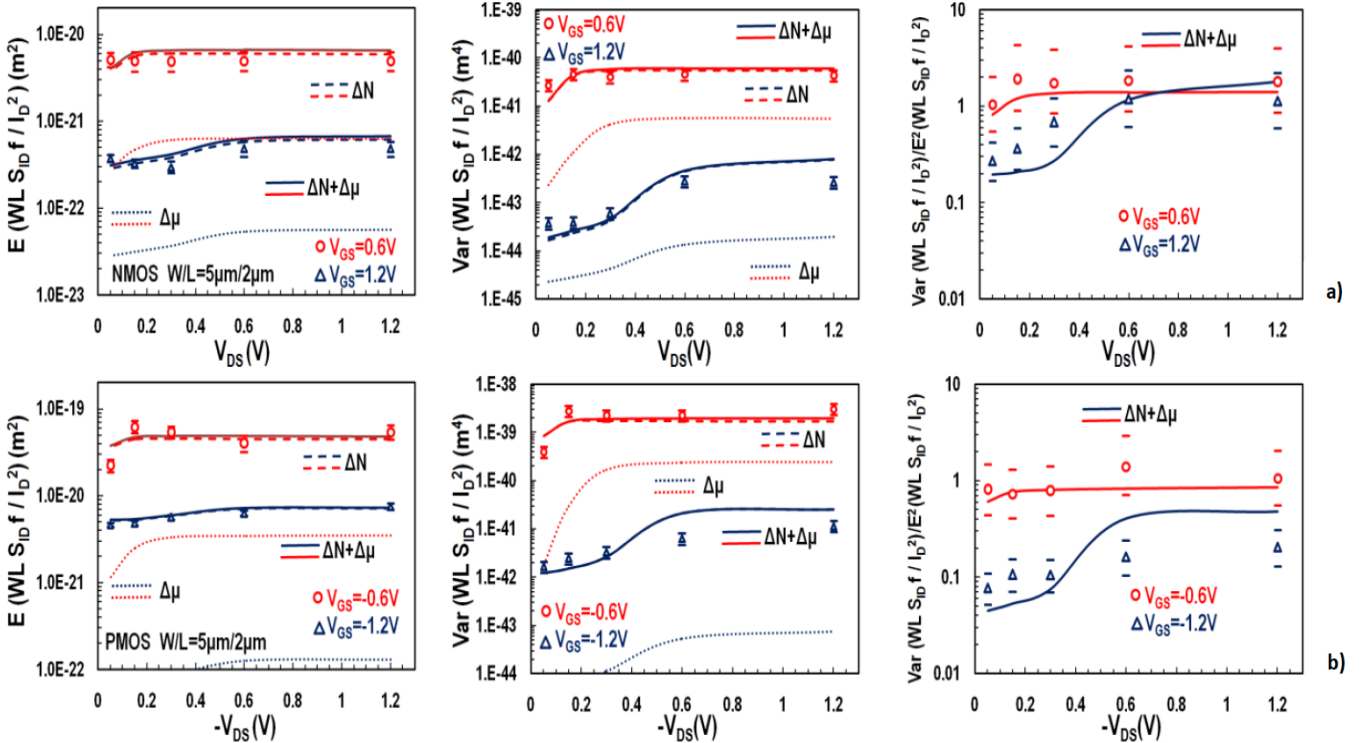

Fig. 4. Mean value (left), variance (center) and normalized variance (right) of output noise $WLS_{ID}/I_D^2$, referred to 1 Hz, vs. drain voltage $V_{DS}$ for both N-channel (a) and P-channel (b) MOSFETs with W=5µm, L=2µm at $|V_{GS}|$=0.6 and 1.2V, respectively. Markers: measured data (with error bars indicating standard error), lines: entire model ($\Delta N + \Delta\mu$), dashed: individual contributions ($\Delta N$, $\Delta\mu$) to 1/f noise.

## V. RESULTS AND DISCUSSION

Bias-dependency of LFN noise statistics is very interesting. As it will be shown below, $\Delta N$ and $\Delta \mu$ effects will determine variance in the region where each effect is dominant in the LFN model. Thus $\Delta N$ will affect 1/f noise variance mostly in moderate to strong inversion, while $\Delta \mu$ in sub-threshold region.

The dashed lines in Fig. 2 represent the $\pm 2\sigma$ standard deviation as obtained from the variance model (18) proposed in the present work, and give a reasonable estimate of the respective data spread (markers). This representation confirms the consistency among physical and statistical charge-based LFN models.

Noise measurements are very sensitive, particularly in weak inversion [17]. As a consequence the handling of statistical noise quantities is very sensitive, particularly so the normalized variance of noise. Based on the log-normal distribution of noise (e.g. [13], [18] – [22]), data in Figs. 3 and 4 show ln-mean values for average noise, while variance is calculated from $\sigma(ln(WLS_{ID}/I_D^2))$, and normalized variance is obtained by inverting the relation $\sigma(ln(WLS_{ID}/I_D^2)) = \sqrt{(ln[1 + Var(WLS_{ID}/I_D^2) / E^2(WLS_{ID}/I_D^2)])}$ [18], [13]. Error bars corresponding to normalized standard error are also shown.

Fig. 3 presents the measured normalized flicker noise, its variance, as well as normalized variance, versus normalized drain current, for both linear and saturation conditions and for either types of devices. The model is shown to correctly represent the average noise $E(WLS_{ID}/I_D^2)$, its variance $Var(WLS_{ID}/I_D^2)$, as well as the normalized variance $Var(WLS_{ID}/I_D^2)/E^2(WLS_{ID}/I_D^2)$. In each case, the markers represent the measured values for all devices on the wafer, while lines represent the statistical noise model integrated in the EKV3 charge-based compact model.

The bias-dependence of the proposed model (18) is analyzed in center graphs of Fig. 3. The variance of normalized output noise at 1Hz – $Var(WLS_{ID}/I_D^2)$ – shows a minimum in strong inversion and it becomes maximum in weak inversion, bearing some similarity with the expected value of noise $E(WLS_{ID}/I_D^2)$ in left graphs of Fig. 3. In addition, in saturation region variance is higher in comparison to linear region, at all levels of drain current [4], [7], [13]. The complete model follows the data qualitatively well and with an appreciable consistency. Dashed lines representing the different noise deviation contributors provide some interesting insight. The $\Delta N$ effect is seen to be dominant in moderate to strong inversion. In weak inversion, the deviations due to $\Delta N$ effect in linear and saturation regions coincide. Furthermore, as the inversion level increases, the correlated number and mobility fluctuation ($\alpha\mu$ product) – most prominent in PMOS – is seen to contribute to statistical LFN variance. This has been mentioned in [4], [7], [16], where however no physics-based compact model has been proposed.

Detailed data is shown for both linear and saturation regimes in weak inversion. The increased deviation under high $V_{DS}$ conditions in comparison to linear region is well modeled through the mobility fluctuation ($\Delta \mu$) effect. The $ln(q_s/q_d)$ term in (17) is dominant in weak inversion and leads to higher variation in saturation, where $(q_s/q_d)$ is large. Conversely, in linear region, $q_d$ is comparable to $q_s$.

Some interesting observations can be made by examining the normalized variance of output noise, $Var(WLS_{ID}/I_D^2)/E^2(WLS_{ID}/I_D^2)$, as shown in the right graphs of Fig. 3. As noted in the Section on $\Delta N$ model, the number of traps $N_{tr} = WLN_t$ is assumed to be Poisson distributed and hence $Var(N_{tr}) = WLN_t = WLkT\lambda N_T$. At low $V_{DS}$, supposing ideally $E_{aC} = 1$, we find that $Var(WLS_{ID}/I_D^2)/E^2(WLS_{ID}/I_D^2) = E_{NT}/(WLN_t)$ independently of bias for the whole range of inversion; see e.g. the discussion in [18], and Fig. 15 and 16 in [19]. Indeed the present $\Delta N(\alpha\mu=0)$ model in linear mode is practically independent of drain current (hence $V_G$) from weak to strong inversion. Other published data show a bias-dependence, e.g. Fig. 11 in [13]. In our model, the bias-dependence of normalized variance occurs when $E_{aC} \neq 1$, because noise variance is affected by $E_{aC}$ while the mean value remains unaffected. In weak or moderate inversion (in absence of $\Delta \mu$ effect), we find the same asymptotic value $E_{NT}/(WLN_t)$, as indicated in right graphs of Fig. 3. For the NMOS case, where $E_{NT} = 1$, the above observation is reasonably well confirmed, if we ignore the incidence of $\Delta \mu$ effect. Hence, this indicates that the above reasoning is correct and supports the Poisson distribution of number of traps $N_{tr}$. For PMOS, the $E_{NT}$ parameter differs from unity, illustrating the usefulness of introducing some flexibility in the statistical model. Accordingly, the above assumption is less well supported in the present PMOS data. In strong inversion saturation, the $\Delta N(\alpha\mu=0)$ model shows increased normalized variance with respect to weak-moderate inversion or linear mode. Normalized variance in strong inversion is reduced for $E_{aC}$ values below unity, corresponding to a reduced impact of the $\alpha\mu$ product in variance. In weak inversion, the $\Delta \mu$ effect leads to an increase in variance, which is enhanced at higher $V_{DS}$. A good estimate for both LFN and its variance at all bias conditions is required for correctly modeling the bias-dependent normalized variance.

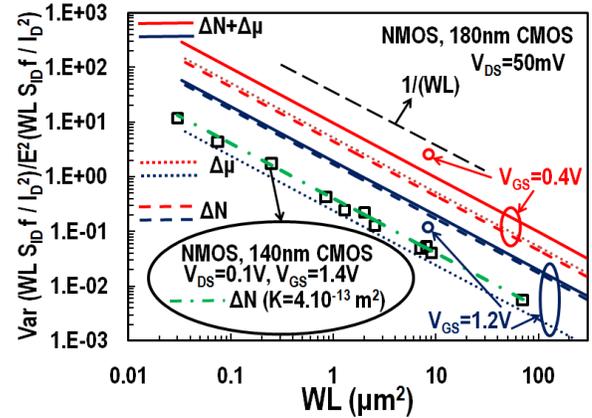

Fig. 5. Normalized variance of normalized output noise vs. area, showing the same ~1/(WL) scaling trends for variability due to $\Delta N$ and $\Delta \mu$ variations. At $V_{GS}=1.2V$, $\Delta N$ effect dominates, while $\Delta \mu$ effect dominates at $V_{GS}=0.4V$ (both at $V_{DS}=50mV$). Lines: entire model ($\Delta N+\Delta \mu$), dashed: $\Delta N$, dotted: $\Delta \mu$ contribution; circles: measured data of 5$\mu$m/2$\mu$m NMOS device (180nm CMOS); squares: NMOS data (140nm CMOS) [13], dash-dotted: $\Delta N$ model for 140nm CMOS, at $V_{GS}=1.4V$, $V_{DS}=0.1V$, confirming the ~1/(WL) scaling trend.

Fig. 4 shows normalized noise, variance, and normalized variance of noise, versus drain voltage, for a gate voltage in moderate and strong inversion, respectively. Both curves are dominated by the $\Delta N$ effect. The same model parameters are used as in Fig. 3. The modeled and the measured behavior concur qualitatively well, given that the model handles linear to saturation behavior without any parameter fitting. The drain voltage dependence of normalized variance is significantly more pronounced in strong inversion. Drain bias dependence of normalized variance can be significantly reduced in short-channel devices [13], which may be due to velocity saturation.

Finally, Fig. 5 illustrates the geometrical scaling inherent in the present statistical LFN model. Either $\Delta N$ and $\Delta \mu$ effects show a scaling of normalized variance vs. area $Var(WLS_{ID}/I_D{}^2)/E^2(WLS_{ID}/I_D{}^2) \sim 1/(WL)$ – or equivalently, a scaling of normalized standard deviation vs. area $\sigma(WLS_{ID}/I_D{}^2)/E(WLS_{ID}/I_D{}^2) \sim 1/\sqrt{(WL)}$, as is observed e.g. in [7] (for short-channel devices), [18]. Measured data from a 140nm CMOS process [13] shown in Fig. 5 fully confirm the $\sim 1/(WL)$ scaling for normalized variance. The present compact model provides, for the first time, analytical expressions for noise variability at all bias conditions. The model's scaling properties are compatible with [13].

## VI. CONCLUSIONS

This paper investigates in detail the bias-dependence of low frequency noise variability in large-area, standard bulk MOSFETs. An analytical statistical compact model is developed, based on number and mobility fluctuation mechanisms $\Delta N$ and $\Delta \mu$, where the LFN variability is obtained from the variability of trap density $N_T$ and Hooge parameter $\alpha_H$, respectively. The resulting compact model allows for the first time to cover analytically the observed variability of LFN over all bias conditions, from weak to strong inversion, as well as from linear to saturation regions. Data measured at our lab is from an experimental 180nm CMOS process on N- and P-channel devices. $\Delta N$ and $\Delta \mu$ effects affect 1/f noise statistics in a consistent and similar way they affect 1/f noise ln-mean value. Typically, $\Delta N$ effect is responsible for noise variance behavior in (weak-) moderate to strong inversion. The examination of normalized variance lends some support for Poisson distributed number of traps $N_{tr}$. On the other hand, the $\Delta \mu$ effect (if present) may dominate in weak-moderate inversion.

This work leads to a consistent model of variance of 1/f noise (18), containing both $\Delta N$ (including correlated $\Delta N$-$\Delta \mu$ noise) (9) and $\Delta \mu$ components (17). The noise variability model is highly consistent with the noise model at all bias conditions. Its parameters can be estimated from the parameters of the noise model itself, and may be further refined based on measured data. The bias-dependence of the statistical 1/f noise model has been validated for both gate and drain bias dependence. The geometrical scaling of the statistical noise model is consistent with data from [13].

## REFERENCES

[1] G. Ghibaudo, O. Roux-dit-Buisson, "Low-frequency fluctuations in scaled-down silicon CMOS devices status and trends", *Europ. Solid-State Device Res. Conf. (ESSDERC)*, pp. 693-700, 1994.

[2] I. Bloom, "1/f noise reduction of metal-oxide-semiconductor transistors by cycling from inversion to accumulation", *Appl. Phys. Lett.* , vol. 58, pp. 1664, 1991.

[3] G. Ghibaudo, O. Roux-dit-Buisson, J. Brini, "Impact of scaling down on low-frequency noise in silicon MOS transistors", *Phys. Stat. Sol. A*, vol. 132, pp. 501-507, 1992.

[4] R. Brederlow, W. Weber, D. Schmitt-Landsiedel, R. Thewes, "Fluctuations of the low frequency noise of MOS transistors and their modeling in analog and RF-circuits", in *IEDM Tech. Dig.*, 1999, pp. 159–162.

[5] G. Ghibaudo, "Electrical noise and RTS fluctuations in advanced CMOS devices", *Microelectronics Reliability*, vol. 42, no. 4-5, pp. 573-582, 2002.

[6] A. P. van der Wel, E. A. M. Klumperinck, E. Hoekstra, B. Nauta, "Relating random telegraph signal noise in metal-oxide-semiconductor transistors to interface trap energy distribution", *Appl. Phys. Lett.* , vol. 87, no. 18, pp. 183507, 2005.

[7] G. I. Wirth, J. Koh, R. Silva, R. Thewes, R. Brederlow, "Modeling of statistical low-frequency noise of deep-submicrometer MOSFETs", *IEEE Trans. Electron Devices*, vol. 52, no. 7, pp. 1576-1588, Jul. 2005.

[8] A. L. McWhorter, "1/f noise and germanium surface properties", in *Semiconductor Surface Physics*, ed. R. H. Kingston, U. Pennsylvania Press, pp. 207-228, 1957.

[9] G. Reimbold, "Modified 1/f trapping noise theory and experiments in MOS transistors biased from weak to strong inversion-influence of interface states", *IEEE Trans. Electron Devices*, vol. 31, no. 9, pp. 1190-1198, Sept. 1984.

[10] K. K. Hung, P. K. Ko, C. Hu, Y. C. Cheng, "A Unified Model for the Flicker Noise in Metal-Oxide-Semiconductor Field-Effect Transistors", *IEEE Trans. Electron Devices*, vol. 37, no. 3, pp. 654-665, Mar. 1990.

[11] G. Ghibaudo, O. Roux, C. Nguyen-Duc, F. Balestra, J. Brini, "Improved analysis of low frequency noise in field-effect MOS transistors", *Phys. Stat. Sol. (a)* vol. 124, no. 2, pp. 571–81, 1991.

[12] A. J. Scholten, L. F. Tiemeijer, R. van Langevelde, R. J. Havens, A. T. A. Zegers-van Duijnhoven, V. C. Venezia, "Noise Modeling for RF CMOS Circuit Simulation", *IEEE Trans. Electron Devices*, vol. 50, no. 3, pp. 618-632, Mar. 2003.

[13] M. Banaszeski da Silva, H. Tuinhout, A. Zegers-van Duijnhoven, G. Wirth, A. Scholten, "A Physics-Based RTN Variability Model for MOSFETs", in *IEDM Tech. Dig.*, pp. 35.2.1–35.2.4, 2014.

[14] F. N. Hooge, "1/f Noise", *Physica*, vol. 83B, pp. 14 - 23, 1976.

[15] L. K. J. Vandamme, F. N. Hooge, "What Do We Certainly Know About 1/f Noise in MOSTs", *IEEE Trans. Electron Devices*, vol. 55, no. 11, pp. 3070-3085, Nov. 2008.

[16] M. Ertürk, T. Xia, W. F. Clark, "Gate Voltage Dependence of MOSFET 1/f Noise Statistics", *IEEE Electron Device Lett.*, vol. 28, no. 9, pp. 812-814, Sept. 2007.

[17] H. Tuinhout, A. Zegers-van Duijnhoven, "Evaluation of 1/f noise variability in the subthreshold region of MOSFETs", *Int. Conf. on Microelectronic Test Structures (ICMTS)*, pp. 87-92, 2013.

[18] T. H. Morshed, M. V. Dunga, J. Zhang, D. D. Lu, A. M. Niknejad, C. Hu, "Compact Modeling of Flicker Noise Variability in Small Size MOSFETs", in *IEDM Tech. Dig.*, pp. 719-722, 2009.

[19] E. G. Ioannidis, S. Haendler, A. Bajolet, T. Pahron, N. Planes, F. Arnaud, R. A. Bianchi, M. Haond, D. Golanski, J. Rosa, C. Fenouillet-Beranger, P. Perreau, C. A. Dimitriadis, G. Ghibaudo, "Low frequency noise variability in high-k/metal gate stack 28nm

bulk and FD-SOI CMOS transistors", in *IEDM Tech. Dig.*, pp. 18.6.1–18.6.4, 2011.

[20] E. G. Ioannidis, S. Haendler, C. G. Theodorou, S. Lasserre, C. A. Dimitriadis, G. Ghibaudo, "Evolution of low frequency noise and noise variability through CMOS bulk technology nodes from 0.5μm down to 20 nm", *Solid-State Electronics*, vol. 95, pp. 28-31, 2014.

[21] E. G. Ioannidis, S. Haendler, A. Bajolet, J. Rossa, J.-P. Manceau, C. A. Dimitriadis, G. Ghibaudo, "Evolution of low frequency noise and noise variability through CMOS bulk technology nodes", *22nd Int. Conf. on Noise and Fluctuations (ICNF)*, pp. 1-4, 2013.

[22] D. Lopez, S. Haendler, C. Leyris, G. Bidal, G. Ghibaudo, "Low-Frequency Noise Investigation and Noise Variability Analysis in High-k/Metal Gate 32-nm CMOS Transistors", *IEEE Trans. Electron Devices*, vol. 58, no. 8, pp. 2310-2316, Aug. 2011.

[23] N. Mavredakis, P. Habas, A. Acovic, R. Meyer, M. Bucher, "Variability of low frequency noise in Moderately-Sized MOSFETs - A Model for the Area- and Gate Voltage- dependence", *23rd Int. Conf. on Noise and Fluctuations (ICNF)*, pp. 1-4, 2015.

[24] A. Arnaud, C. Galup-Montoro, "A compact model for flicker noise in MOS Transistors for Analog Circuit Design", *IEEE Trans. Electron Devices*, vol. 50, no. 8, pp. 1815-1818, Jul. 2003.

[25] C. Enz, E. Vittoz, Charge-based MOS Transistor Modeling, *John Wiley and Sons*, Chichester, 2006.

[26] A. S. Roy, C. C. Enz, "Critical Discussion of the Flatband Perturbation Technique for Calculating Low-Frequency Noise", *IEEE Tans. Electron Devices*, vol. 53, no. 10, pp. 2664-2667, Oct. 2006.

[27] N. Mavredakis, A. Antonopoulos, M. Bucher, "Measurement and Modeling of 1/f Noise in 180 nm NMOS and PMOS Devices", *5th Europ. Conf. on Circuits & Systems for Communications (ECCSC)*, pp. 86-89, 2010.

[28] J.-M. Sallese, M. Bucher, F. Krummenacher, P. Fazan, "Inversion charge linearization in MOSFET modeling and rigorous derivation of the EKV compact model", *Solid-State Electronics*, vol. 47, no. 4, pp. 677–683, Apr. 2003.